\documentclass[12pt,a4paper]{article}

\usepackage[utf8]{inputenc}
\usepackage{graphicx}
\usepackage{amsmath}
\usepackage{amssymb}
\usepackage{booktabs}
\usepackage{array}
\usepackage{multirow}
\usepackage{xcolor}
\usepackage[numbers,sort&compress]{natbib}
\usepackage{setspace}
\usepackage[margin=2.5cm]{geometry}
\usepackage{xurl}
\usepackage{float}
\usepackage{placeins}
\usepackage{caption}
\usepackage{hyperref}
\usepackage{tikz}
\usetikzlibrary{shapes.geometric, arrows.meta, positioning, fit}

\definecolor{pastelBrown}{RGB}{195,176,145}
\definecolor{pastelBlue}{RGB}{174,214,241}
\definecolor{pastelRed}{RGB}{244,177,162}
\definecolor{pastelTeal}{RGB}{162,217,206}
\definecolor{pastelNavy}{RGB}{174,182,191}
\definecolor{paperBlue}{RGB}{25,72,106}

\hypersetup{
    colorlinks=true,
    linkcolor=paperBlue,
    citecolor=paperBlue,
    urlcolor=paperBlue
}

\begin{document}

\title{EEGPrep: a validated Python implementation of the EEGLAB preprocessing pipeline}

\author{
Arnaud Delorme$^{1,2,*}$, Suraj Ranganath$^{1}$, Christian Kothe$^{3}$, \\
Aman Jaiswal$^{1}$, Scott Makeig$^{1}$, Bruno Aristimunha$^{1,4}$ \\
\\
\small $^1$Swartz Center for Computational Neuroscience, University of California San Diego, \\
\small La Jolla, CA, USA \\
\small $^2$Centre de Recherche Cerveau et Cognition, CNRS, Toulouse, France \\
\small $^3$Intheon, La Jolla, CA, USA \\
\small $^4$Yneuro, Paris, France \\
\\
\small $^{*}$Author to whom any correspondence should be addressed: adelorme@ucsd.edu \\
}

\date{}

\maketitle

\begin{abstract}
\begin{singlespace}
\noindent \textbf{Objective.} 
Automated EEG preprocessing is common in research and clinical work, but few pipelines have been tested systematically. In a recent benchmark, the default EEGLAB pipeline was the only pipeline that significantly outperformed simple high-pass filtering. Its MATLAB implementation, however, complicates deployment in Python and cloud workflows. We developed EEGPrep to reproduce this pipeline in Python while supporting BIDS data. The main technical problem was numerical: small floating-point differences can accumulate during recursive filtering and ICA.

\noindent \textbf{Approach.}
EEGPrep covers the default EEGLAB workflow: average re-reference, artifact removal with the clean\_rawdata plugin, Picard ICA, ICLabel plugin component classification, channel interpolation, and epoching. We compared each stage with MATLAB reference output on ARM arm64 (primary analysis) and Intel x86\_64 (supplementary analysis). The test dataset contained 64-channel P300 auditory oddball recordings from 13 participants. We measured maximum absolute difference, RMS error, AMARI distance for ICA, and agreement between ICLabel decisions.

\noindent \textbf{Main Results.}
On ARM arm64, import and re-reference matched exactly for all 12 analysed subjects. The clean\_rawdata and Picard ICA stages remained at numerical zero (maximum RMS~$= 1.5 \times 10^{-12}$~µV; AMARI distance~$\leq 0.000001$; mean correlation~$= 1.000$). ICLabel neural-network inference introduced the only measurable difference (maximum RMS~$= 2.0 \times 10^{-5}$~µV). Rejection decisions nevertheless agreed for every subject, and the difference did not increase through interpolation and epoching (end-to-end maximum RMS~$\leq 2.1 \times 10^{-5}$~µV). The Intel x86\_64 analysis matched to the same precision.

\noindent \textbf{Significance.}
EEGPrep reproduces the EEGLAB validated pipeline and can be installed from PyPI or run in Docker. In addition to processing individual subjects, the full workflow may read BIDS input and write BIDS derivatives. 
\end{singlespace}
\end{abstract}

\begin{singlespace}
\end{singlespace}

\section{Introduction}

Many EEG studies now depend on automated preprocessing, yet most pipelines have not been tested systematically against established standards. Researchers therefore have little basis for deciding whether two pipelines differ because of their algorithms, their implementations, or both. This problem sits within the wider reproducibility concerns documented in neuroscience \citep{button2013power}.

MATLAB remains central to EEG analysis, but licensing and containerization make MATLAB workflows harder to run on shared or cloud infrastructure. Python is already supported by common data-science and cloud platforms. A Python port of an established workflow is useful only if its behavior can be checked against the reference, rather than assumed from algorithm names alone.

EEGLAB \citep{delorme2004eeglab} is an open-source EEG toolbox with more than 20,000 citations and users in hundreds of laboratories. A recent benchmark compared four pipelines (EEGLAB, FieldTrip, MNE-Python, and Brainstorm) on three public datasets after scanning and optimizing their parameters using a cross validation scheme \citep{delorme2023eeg}. Several other published EEGLAB pipelines were also considered, including HAPPE \citep{gabard2018happe} and PREP \citep{bigdely2015prep}, which includes a noise-resistant reference procedure. After parametrization, only the standard EEGLAB pipeline performed significantly better than simple high-pass filtering, increasing detection of experimental effects by 5--18\%. The authors attributed much of this gain to the bad-channel detection and interpolation in the clean\_rawdata EEGLAB plugin. While, the EEGLAB, FieldTrip, and Brainstorm pipelines were evaluated in MATLAB, the MNE-Python pipeline used Autoreject \citep{jas2017autoreject}, and the configuration assessed in the benchmark did not compare favorably with the other tested pipelines \citep{delorme2023eeg}.

The default EEGLAB workflow uses the clean\_rawdata plugin \citep{mullen2015real} for artifact rejection and the ICLabel plugin \citep{pion2019iclabel} for ICA component classification. Within clean\_rawdata, Artifact Subspace Reconstruction (ASR) identifies high-amplitude transients while retaining the remaining signal \citep{mullen2015real}. ICLabel assigns each component to brain, muscle, eye, heart, line noise, channel noise, or other. Its labels can then drive component rejection without manual review. Previous work on ICLabel found that, after careful optimization, Python and MATLAB implementations agreed within 0.001\% \citep{delorme2024iclabel}.

Exact agreement across languages is difficult. Floating-point arithmetic is not associative, and math libraries may use different reduction orders \citep{goldberg1991every}. Recursive and iterative computations can magnify the last-bit discrepancies: filters feed their output back as input, while an ICA solver may follow another optimization path after a small perturbation. Related differences have been reported across neuroimaging packages \citep{bowring2019exploring}. 

This study asks whether a Python implementation can reproduce the default EEGLAB preprocessing workflow and, when it cannot, where the first numerical difference appears. We implemented clean\_rawdata, Picard ICA, and ICLabel within the same seven-step pipeline, then compared each intermediate output with MATLAB. A separate analysis measured ICA variability to put cross-language differences in context. After more than one year of careful optimization, here we present a numerically equivalent port of the EEGLAB MATLAB pipeline to Python.

\section{Methods}

\subsection{Pipeline architecture}

\begin{figure}[!t]
\centering
\begin{tikzpicture}[
    scale=0.85, transform shape,
    node distance=0.65cm,
    box/.style={rectangle, rounded corners, minimum width=4.5cm, minimum height=0.7cm,
                text centered, draw=black, line width=0.8pt, font=\small},
    tallbox/.style={box, minimum height=1.0cm},
    arrow/.style={-{Stealth[length=2.5mm]}, line width=1.0pt},
    ]

    \node[box, fill=pastelBrown, font=\small\bfseries] (input) {BIDS Dataset (Input)};

    \node[box, fill=pastelBlue, below=of input] (s1) {Step 1: Import};

    \node[box, fill=pastelBlue, below=of s1] (s2) {Step 2: Re-reference};

    \node[box, fill=pastelRed, below=of s2, font=\small\bfseries] (s3) {Step 3: clean\_rawdata};

    \node[box, fill=pastelTeal, below=of s3, font=\small\bfseries] (s4) {Step 4: Picard ICA};

    \node[box, fill=pastelTeal, below=of s4, font=\small\bfseries] (s5) {Step 5: ICLabel};

    \node[box, fill=pastelBlue, below=of s5] (s6) {Step 6: Interpolation};

    \node[box, fill=pastelNavy, below=of s6] (s7) {Step 7: Epoching};

    \node[box, fill=pastelBrown, below=of s7, font=\small\bfseries] (output) {BIDS Derivatives (Output)};

    \draw[arrow] (input) -- (s1);
    \draw[arrow] (s1) -- (s2);
    \draw[arrow] (s2) -- (s3);
    \draw[arrow] (s3) -- (s4);
    \draw[arrow] (s4) -- (s5);
    \draw[arrow] (s5) -- (s6);
    \draw[arrow] (s6) -- (s7);
    \draw[arrow] (s7) -- (output);

    \node[box, fill=pastelBrown, minimum width=1.0cm, minimum height=0.4cm, font=\tiny]
        at (4.5, -0.8) {};
    \node[anchor=west, font=\footnotesize] at (5.2, -0.8) {I/O};

    \node[box, fill=pastelBlue, minimum width=1.0cm, minimum height=0.4cm, font=\tiny]
        at (4.5, -1.5) {};
    \node[anchor=west, font=\footnotesize] at (5.2, -1.5) {Preprocessing};

    \node[box, fill=pastelRed, minimum width=1.0cm, minimum height=0.4cm, font=\tiny]
        at (4.5, -2.2) {};
    \node[anchor=west, font=\footnotesize] at (5.2, -2.2) {Artifact Removal};

    \node[box, fill=pastelTeal, minimum width=1.0cm, minimum height=0.4cm, font=\tiny]
        at (4.5, -2.9) {};
    \node[anchor=west, font=\footnotesize] at (5.2, -2.9) {ICA/ICLabel};

    \node[box, fill=pastelNavy, minimum width=1.0cm, minimum height=0.4cm, font=\tiny]
        at (4.5, -3.6) {};
    \node[anchor=west, font=\footnotesize] at (5.2, -3.6) {Epoching};

\end{tikzpicture}
\caption{The seven-step EEGPrep workflow from BIDS input to BIDS derivatives. The steps match those in the parity analysis (Table~\ref{tab:parity_summary}). Colours identify I/O (brown), preprocessing (blue), artifact removal (red), ICA/ICLabel (teal), and epoching (navy). Intermediate results can be saved at every step.}
\label{fig:pipeline}
\end{figure}
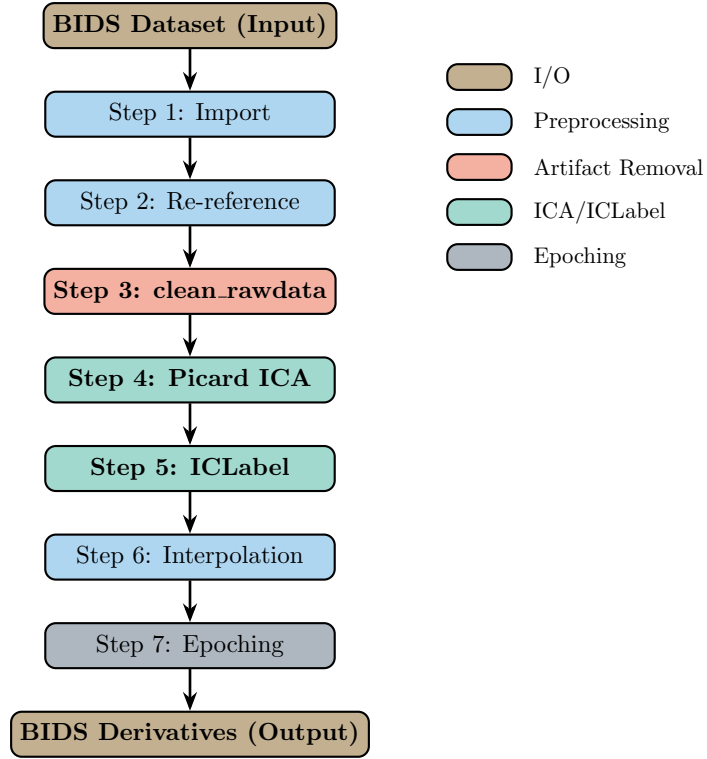

EEGPrep follows the default EEGLAB workflow in seven steps (Figure~\ref{fig:pipeline}), which correspond to the stages in the parity analysis (Table~\ref{tab:parity_summary}). It reads BIDS-formatted EEG data and writes a BIDS derivatives dataset.

\begin{itemize}
    \item \noindent Step~1 (\emph{Import}). EEG data are loaded from the BIDS dataset. Channels classified as EOG, EMG, or triggers are excluded.
    \item \noindent Step~2 (\emph{Re-reference}). Signals are re-referenced to the common average.
    \item \noindent Step~3 uses \emph{clean\_rawdata} to remove bad channels, high-amplitude transients, and noisy time periods \citep{mullen2015real}. Artifact Subspace Reconstruction (ASR) uses default parameters calibrated on a clean reference period.
    \item \noindent Step~4 (\emph{Picard ICA}). Independent component analysis is performed with Picard \citep{ablin2018faster}.
    \item \noindent Step~5 (\emph{ICLabel}). ICLabel classifies each component as brain, muscle, eye, heart, line noise, channel noise, or other \citep{pion2019iclabel}. Components assigned to muscle or eye with probability $\geq 0.9$ are removed, following the default EEGLAB optimized thresholds.
    \item \noindent Step~6 (\emph{Interpolation}). Spherical-spline interpolation reconstructs channels removed by clean\_rawdata and restores the full montage.
    \item \noindent Step~7 (\emph{Epoching}). The continuous recording is segmented around experimental events.
\end{itemize}

The pipeline can save each intermediate result in EEGLAB \texttt{.set} format for direct comparison. Channel locations, events, and experimental parameters are retained.

\subsection{Implementation details}

EEGPrep is written for Python 3.8 and later. It uses NumPy for numerical operations \citep{harris2020array} and SciPy for signal processing \citep{virtanen2020scipy}. Picard optimizes the same maximum-likelihood (Infomax) objective as EEGLAB's default \texttt{runica} \citep{bell1995information}, but uses a deterministic quasi-Newton solver \citep{ablin2018faster}. Its fixed convergence path permits an exact cross-language comparison.

The earlier MATLAB port of Picard contained several minor discrepancies from the canonical Python implementation maintained by the algorithm's authors. We corrected the MATLAB port used for the reference output, bringing the two implementations into exact agreement. ICLabel inference runs in PyTorch \citep{paszke2019pytorch} with weights ported directly from the MATLAB model.

The Python EEG data structure is implemented as a dictionary that mirrors the MATLAB EEGLAB structure field by field. The Python functions retain the names of their MATLAB counterparts, while their parameters are adapted to Python conventions. 

\smallskip

\noindent\textit{BIDS integration.} The core functions operate on individual EEG files and do not require a dataset layout. For this validation, we placed an optional BIDS layer above them. Given a dataset that follows BIDS for EEG \citep{pernet2019eeg}, this layer locates the EEG, channel-location, and event files from their names. It writes the BIDS output and metadata, including \texttt{dataset\_description.json} and \texttt{participants.tsv}, to a BIDS derivatives directory.

\smallskip

\noindent\textit{Random-seed control.} Several steps use random samples, especially the RANSAC bad-channel detection in clean\_rawdata. Each stochastic function accepts a seed. NumPy's \texttt{numpy.random.RandomState} and MATLAB's default \texttt{rng(\ldots,\,'twister')} share the MT19937 core. With the MATLAB default seed of 5489, they produce the same uniform random sequence. However, the Gaussian random draws use a different algorithm in Python and MATLAB, so EEGPrep builds every stochastic operation from uniform draws. RANSAC channel sampling and permutation use a swap-based Fisher--Yates shuffle, consume uniform random values, and apply MATLAB's round-half-away-from-zero rule. The resulting stochastic decisions are thus identical.

\smallskip

\noindent\textit{Distribution.} The PyPI package can be installed with \texttt{pip install eegprep} and Docker images are also provided. The smaller installation omits PyTorch and stops at clean\_rawdata. The full installation adds ICLabel and occupies approximately 2~GB because of the PyTorch dependency. An EEGPrep container was deployed on the brainlife.io cloud computing platform (\url{https://brainlife.io/app/6793e2ee81d348aa5654e893}).

\subsection{Testing framework}

We compared EEGPrep with EEGLAB 2026.0.0. To evaluate individual stages, both pipelines first receive byte-identical input. They then receive the output from the previous stage in their respective pipelines to assess cumulative pipeline drift.

The source was the BIDS dataset NEMAR ds003061 (\url{https://nemar.org}), a P300 auditory oddball task with 70\% standard, 15\% oddball, and 15\% distractor trials. The dataset contains 64-channel BioSemi Active~2 recordings from 13 participants, sampled at 256~Hz. We analysed run~1. Participant sub-012 was excluded because only approximately 10\% of the expected stimulus events were present, and no data remained for epoching after artefact rejection. The final analysis therefore included 12 recordings.

The primary analysis ran on ARM arm64 (macOS, Apple Silicon) with Python 3.11 and MATLAB R2025a (Apple Accelerate). We repeated it on Intel x86\_64 (macOS) with Python 3.10 and MATLAB R2022b (Intel MKL), with results reported in Supplementary Table~\ref{tab:intel_parity_summary}. On each architecture, MATLAB ran through the MATLAB Engine and the Python pipeline used its native PyTorch ICLabel implementation. The Intel run required NumPy 1.x. That numerical configuration produces small differences from the ARM reference, as discussed in the Discussion section.

The pytest suite invokes both implementations with the same inputs and parameters, using a compatibility layer to call the original, unmodified EEGLAB functions through the MATLAB Engine for Python. This enables direct comparison at every stage.

\subsection{Comparison metrics}

For each processing stage, we computed the following metrics to quantify differences between Python and MATLAB implementations:
\begin{itemize}
\item \textbf{Maximum absolute difference:} $\max(|X_{py} - X_{mat}|)$ across all data points
\item \textbf{Mean absolute difference:} $\text{mean}(|X_{py} - X_{mat}|)$
\item \textbf{Root mean square (RMS):} $\sqrt{\text{mean}((X_{py} - X_{mat})^2)}$
\end{itemize}

where $X_{py}$ and $X_{mat}$ are the Python and MATLAB EEG data matrices at a given stage (channels $\times$ time points).

Cumulative metrics measure the total difference from the initial import stage through the current stage and capture error propagation across the pipeline. Incremental metrics isolate the contribution of each stage by computing differences relative to the previous stage.

For Picard ICA at Step~4, we matched Python and MATLAB components by applying the Hungarian algorithm to the absolute cross-correlation matrix \citep{kuhn1955hungarian}. This bipartite assignment maximizes the sum of correlations across all matched pairs. Matching used columns of the mixing matrix, which represent scalp projections.

For each matched component pair $i$, we computed Pearson correlation:
\begin{equation}
r_i = \frac{\text{cov}(IC_{py,i}, IC_{mat,i})}{\sigma(IC_{py,i}) \cdot \sigma(IC_{mat,i})}
\end{equation}

We report the mean, minimum, and maximum correlation across components, as well as the AMARI distance \citep{amari1996new}.

After Step~5, we compared the ICLabel probabilities, component labels, and rejection decisions produced by the two implementations.

\section{Results}

\subsection{Stage-by-stage numerical accuracy}

\begin{figure}[!t]
\centering
\includegraphics[width=\textwidth]{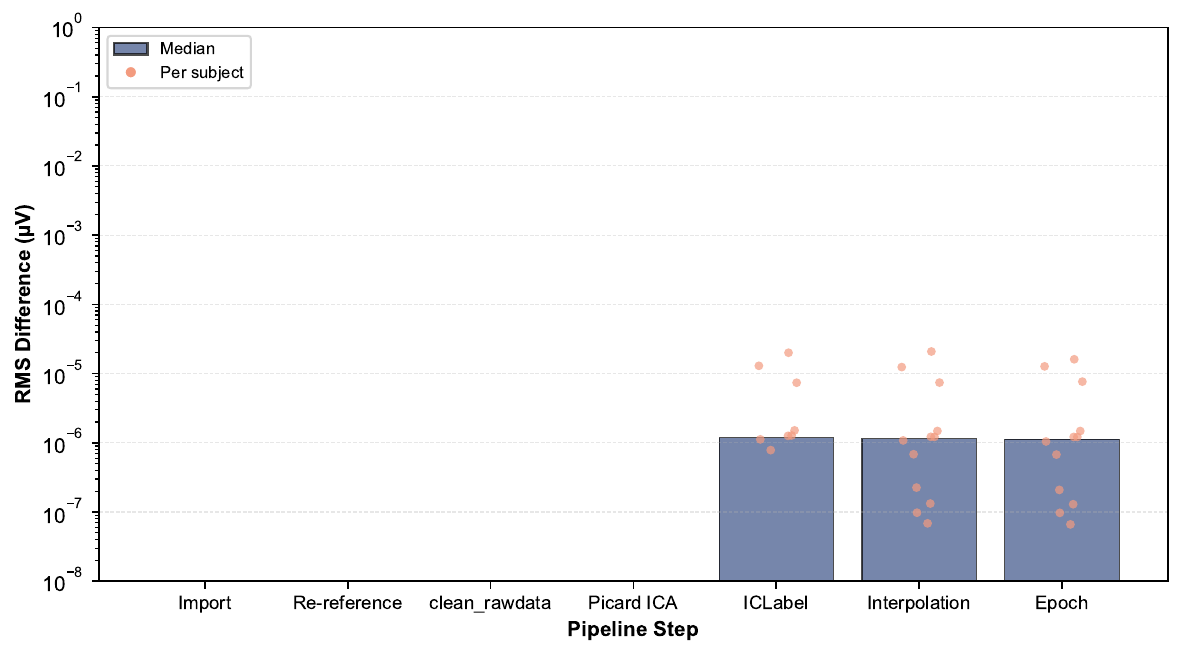}
\caption{RMS difference between Python EEGPrep and MATLAB at each pipeline step (run~1, ARM arm64, 12 subjects). Bars show medians and dots show subjects. RMS is zero or approximately $10^{-12}$~µV through Step~4. The first non-zero error occurs during ICLabel inference at Step~5 and changes little during interpolation and epoching.}
\label{fig:error_propagation}
\end{figure}

The primary ARM arm64 analysis included 12 P300 recordings (64-channel BioSemi Active~2, 256~Hz, run~1; sub-012 excluded). We compared Python and MATLAB output after each of the seven stages using maximum absolute difference and RMS in microvolts (Table~\ref{tab:parity_summary}). Figure~\ref{fig:error_propagation} shows each subject's RMS error at every step. Supplementary Table~\ref{tab:intel_parity_summary} reports the Intel x86\_64 results.

\begin{table}[h]
\centering
\caption{Stage-by-stage numerical parity between Python eegprep and MATLAB eeglabcompat pipelines (run~1). MaxAbsDiff and RMS computed across all channels and time points. Values in µV.}
\label{tab:parity_summary}
\begin{tabular}{lrccc}
\toprule
\textbf{Step} & $N$ & \textbf{Mean RMS} & \textbf{Max RMS} & \textbf{Max MaxAbsDiff} \\
\midrule
1. Import & 12 & 0 & 0 & 0 \\
2. Re-reference & 12 & 0 & 0 & 0 \\
3. clean\_rawdata & 12 & $1.0 \times 10^{-12}$ & $1.5 \times 10^{-12}$ & $1.8 \times 10^{-11}$ \\
4. Picard ICA & 12 & $1.0 \times 10^{-12}$ & $1.5 \times 10^{-12}$ & $1.8 \times 10^{-11}$ \\
5. ICLabel & 12 & $3.9 \times 10^{-6}$ & $2.0 \times 10^{-5}$ & $4.7 \times 10^{-4}$ \\
6. Interpolation & 12 & $3.9 \times 10^{-6}$ & $2.1 \times 10^{-5}$ & $4.7 \times 10^{-4}$ \\
7. Epoch & 12 & $3.5 \times 10^{-6}$ & $1.6 \times 10^{-5}$ & $2.3 \times 10^{-4}$ \\
\bottomrule
\end{tabular}
\end{table}

Import and re-reference produced zero difference across every channel and time point. At clean\_rawdata, both implementations rejected the same channels and time points for all 12 subjects. Output matched to numerical precision through Picard ICA, with RMS values of approximately $10^{-12}$~µV (Table~\ref{tab:parity_summary}).

The first measurable difference appeared at ICLabel (Step~5): mean RMS was $3.9 \times 10^{-6}$~µV (maximum $2.0 \times 10^{-5}$), and maximum absolute difference reached $4.7 \times 10^{-4}$~µV. PyTorch and MATLAB produce slightly different float32 classification probabilities during neural-network inference. Even so, the rejection decisions matched for all 12 subjects.

\begin{figure}[!t]
\centering
\includegraphics[width=\textwidth]{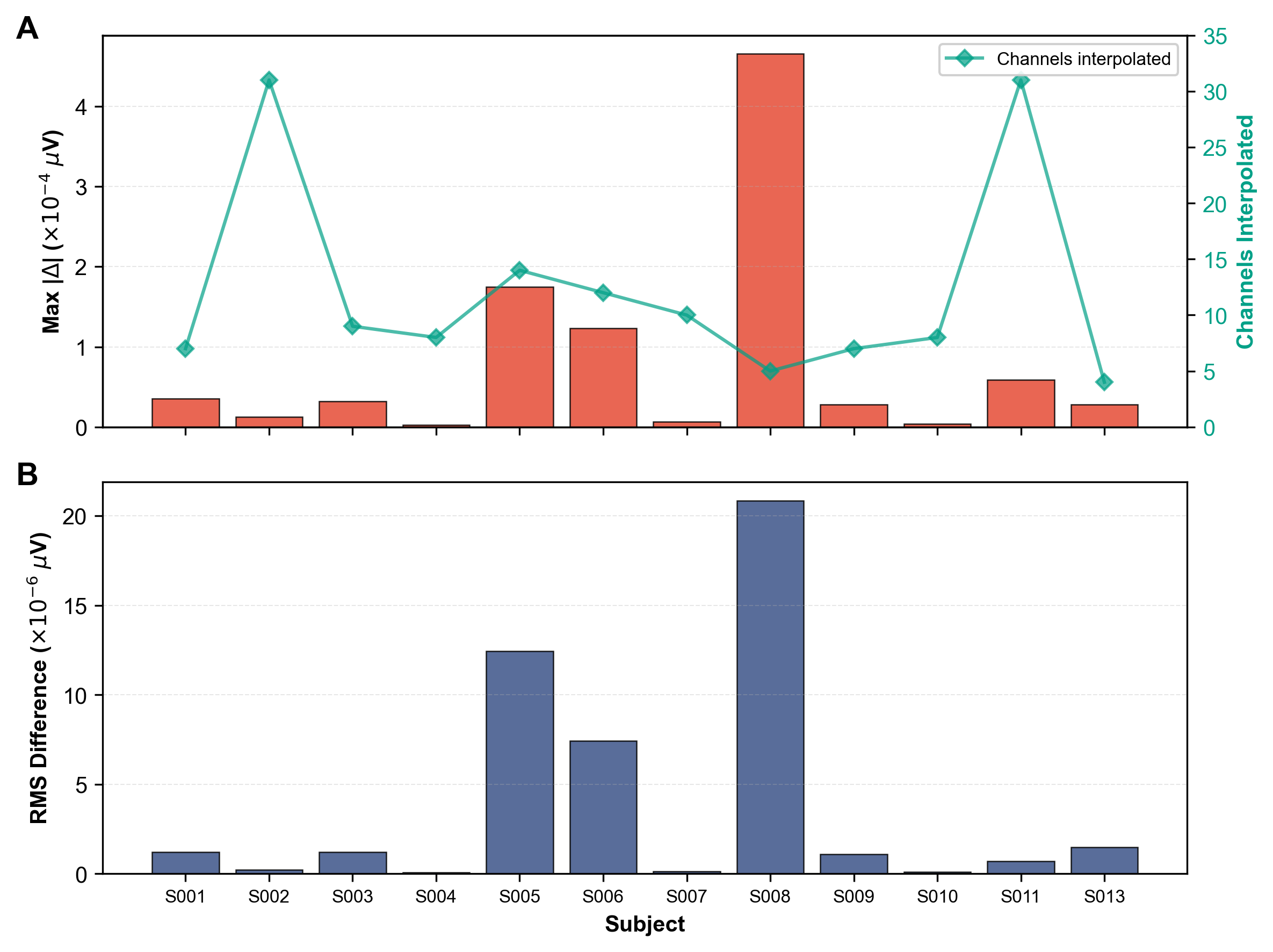}
\caption{Parity after interpolation for 12 ARM arm64 subjects (run~1). Step~6 reconstructs channels removed by clean\_rawdata and restores the 64-channel montage. (\textbf{A}) Maximum absolute difference in the continuous data; green diamonds show the number of interpolated channels. (\textbf{B}) RMS difference across channels and time points. Maximum absolute difference was $\leq 4.7 \times 10^{-4}$~µV and RMS was $\leq 2.1 \times 10^{-5}$~µV. The error enters at ICLabel and does not increase with interpolation or with the number of reconstructed channels.}
\label{fig:interpolation_parity}
\end{figure}

At Step~6, spherical-spline interpolation restored the removed channels. The incoming difference did not measurably increase: mean RMS was $3.9 \times 10^{-6}$~µV and the maximum was $2.1 \times 10^{-5}$~µV (Figure~\ref{fig:interpolation_parity}). Nor did it vary with the number of interpolated channels. After epoching, mean RMS was $3.5 \times 10^{-6}$~µV (maximum $1.6 \times 10^{-5}$), so every end-to-end difference remained below one microvolt.

\FloatBarrier

\subsection{clean\_rawdata rejection parity}

Python and MATLAB clean\_rawdata rejected the same channels and time points for every subject (Table~\ref{tab:asr_decisions}). The number of rejected channels ranged from 4 for sub-013 to 31 for sub-002 and sub-011, out of 64. With the same input, the two implementations therefore made the same discrete rejection decisions.

\begin{table}[!htbp]
\centering
\caption{clean\_rawdata rejection decisions per subject (run~1, ARM arm64). Channel and time-point counts were identical between Python and MATLAB for all 12 analysed subjects (sub-012 excluded).}
\label{tab:asr_decisions}
\begin{tabular}{lccc}
\toprule
\textbf{Subject} & \textbf{Retained Ch.} & \textbf{Rejected Ch.} & \textbf{Time Points} \\
\midrule
sub-001 & 57 & 7 & 187,971 \\
sub-002 & 33 & 31 & 149,647 \\
sub-003 & 55 & 9 & 185,619 \\
sub-004 & 56 & 8 & 169,692 \\
sub-005 & 50 & 14 & 191,669 \\
sub-006 & 52 & 12 & 195,543 \\
sub-007 & 54 & 10 & 121,668 \\
sub-008 & 59 & 5 & 182,255 \\
sub-009 & 57 & 7 & 154,286 \\
sub-010 & 56 & 8 & 181,548 \\
sub-011 & 33 & 31 & 176,640 \\
sub-013 & 60 & 4 & 166,132 \\
\bottomrule
\end{tabular}
\end{table}

\FloatBarrier

\subsection{ICA decomposition fidelity}

Both implementations ran Picard with the same parameters \citep{ablin2018faster}. For every subject, the AMARI distance was $\leq 0.000001$ and mean component correlation was 1.000. Supplementary Table~\ref{tab:ica_parity_json} lists the values by subject.

Picard is a deterministic quasi-Newton method with identity initialization and SVD sign normalization. It therefore avoids the seed-dependent variability of Infomax ICA. Given the same input, the Python and MATLAB implementations followed the same numerical path for all 12 subjects.

\subsection{ICLabel classification comparison}

The two ICLabel implementations made the same rejection decision for every subject (Supplementary Table~\ref{tab:iclabel_parity}). Four subjects had no rejected components; the largest count was three for sub-005. In total, both implementations rejected 15 components. Small float32 differences in PyTorch and MATLAB inference produced a maximum RMS of $2.0 \times 10^{-5}$~µV and a maximum absolute difference of $4.7 \times 10^{-4}$~µV, but none changed a decision.

\subsection{Computational performance}

On a 2.4~GHz eight-core system with 32~GB RAM, the complete pipeline processed sub-001 in approximately 8.5 minutes with Python and 9.2 minutes with MATLAB. Picard accounted for about 60\% of each run. Memory use peaked near 4~GB during ICA.

\section{Discussion}

Across 12 P300 recordings, Python and MATLAB matched to numerical precision through Picard ICA on ARM arm64. The Intel x86\_64 analysis gave the same result. At Step~3, both implementations rejected the same channels and time points and the Picard AMARI distance did not exceed 0.000001. The first measurable difference appeared during ICLabel inference. Within each architecture, it did not change a rejection decision or grow during interpolation and epoching. The largest RMS difference was $2.1 \times 10^{-5}$~µV.

\subsection{Numerical precision and its limits}

The stage-level comparison locates the source of each difference. Import and re-reference matched exactly. After Steps~3 and 4, RMS was approximately $10^{-12}$~µV, which is numerical zero at the scale of these data.

Both implementations calculate the average reference as $(\mathbf{I} - \mathbf{J}/n) \cdot \mathbf{X}$. On ARM, NumPy used Apple Accelerate and MATLAB R2025a used MATLAB's ARM libraries, yet their output still agreed through Picard. On Intel, both programs linked Intel MKL and therefore used the same accumulation order (Supplementary Table~\ref{tab:intel_parity_summary}). Reaching this agreement required several corrections to the Python port. One was replacing \texttt{data - mean(data)} with the matrix multiplication used by MATLAB's \texttt{reref.m}.

Last-bit agreement still depends on compatible floating-point reductions. Re-reference and clean\_rawdata use matrix multiplications, so their output depends on the BLAS accumulation order. The two stacks agreed on both tested architectures, but this need not hold for another library build.

We tested this dependence by forcing NumPy to use OpenBLAS, as required for PyTorch ICLabel with an older NumPy version on Intel macOS. Re-reference then differed by approximately $5 \times 10^{-3}$~µV at the first matrix operation. Picard amplified this float32-ULP-scale perturbation to as much as $0.15$~µV RMS in the epoched output, and ASR burst boundaries could move by a few samples. Channel-rejection and ICLabel decisions did not change in this test. Exact replication should therefore report the random seed and BLAS build; last-bit identity should be expected only with compatible backends. The observed residuals are physiologically negligible, but ICA can amplify them in other environments.

ICLabel introduced the only measurable error on ARM, with a maximum RMS of $2.0 \times 10^{-5}$~µV. The source is float32 inference in PyTorch and MATLAB, which yielded slightly different probabilities but the same decisions. Spherical-spline interpolation neither increased the error nor made it depend on the number of reconstructed channels (Figure~\ref{fig:interpolation_parity}). After epoching, RMS remained below $1.6 \times 10^{-5}$~µV for every subject.

\subsection{Implications for reproducibility}

The relevant scientific question is whether an implementation difference can alter a physiological or statistical conclusion. Exact equality is useful for debugging, but it is not the final validation target.

Here, final RMS differences below $2 \times 10^{-5}$~µV are 6--7 orders of magnitude smaller than typical EEG amplitudes: 10--100~µV for event-related potentials and approximately 5~µV for resting-state oscillations. Measurable error came only from ICLabel inference. For comparison, thermal noise is about 0.1~µV and quantization noise is 0.0061~µV for a 16-bit system with a $\pm$200~µV range and environmental interference adds further variation. Session-to-session EEG correlations are typically 0.6--0.9 \citep{cassidy2012retest}. The implementation differences measured here are much smaller than any of these sources.

These measurements suggest four practical precautions for exact reproducibility:
\begin{enumerate}
\item Report the random seed and the linear-algebra library build. A seed controls RANSAC sampling in clean\_rawdata, while the BLAS backend controls reduction order.
\item Process conditions that will be compared directly, such as drug and placebo recordings, on the same platform and software version.
\item Treat effects near the measured platform variation with caution. An effect below 1~µV may be statistically significant without being physiologically meaningful.
\item Test preprocessing on more than one dataset and compare downstream measures such as signal-to-noise ratio and effect size.
\end{enumerate}

After comparing the data matrices, one should check whether any differences affect the experimental results. A difference matters only if it changes whether an effect is detected, and small differences between two arrays do not matter by themselves.

\subsection{Comparison with other pipelines and tools}

Delorme's benchmark \citep{delorme2023eeg} included MNE-Python's Autoreject \citep{gramfort2013meg,jas2017autoreject}, artifact rejection from FieldTrip and Brainstorm, PREP's robust-reference procedure \citep{bigdely2015prep}, and the EEGLAB-based HAPPE pipeline \citep{gabard2018happe}. The EEGLAB pipeline implemented here was the only one to perform significantly better than high-pass filtering \citep{delorme2023eeg}. Autoreject, for example, rejected 19\% of Oddball trials at $p<0.0001$, but the surviving trials had less statistical power. PREP and other re-reference procedures also reduced the percentage of significant channels.

Other methods fell outside that benchmark. Recent artifact-removal approaches such as ARMBR \citep{alkhoury2025artifact}, which uses multivariate backward regression for blink removal, and ICA with continuous wavelet transform \citep{maddirala2022ica} for low-channel systems, target specific artifact types rather than complete preprocessing workflows and have not undergone comparable effect-level benchmarking.

\subsection{Cloud deployment and BIDS integration}

EEGPrep can be installed from PyPI without a MATLAB licence or manual toolbox setup. It runs in common Python environments, including Google Colab, AWS SageMaker, and Azure ML. The Docker image uses fixed versions of all dependencies and numerical libraries, including the BLAS backend that affected the comparisons above. Each subject and run is processed independently, allowing a cluster scheduler to distribute the tasks.

The optional BIDS layer provides interoperability throughout the workflow. It reads and writes data in BIDS format and can process more than 500 public MEEG datasets accessible through NEMAR \citep{delorme2022nemar}. To our knowledge, EEGPrep is the only pipeline that produces derived BIDS data that pass the BIDS Validator, rather than merely writing output to a derivatives folder, and includes the metadata required by downstream BIDS tools.

EEGPrep can also be run natively on brainlife.io \citep{hayashi2024brainlife}, which records processing provenance, and the Neuroscience Gateway \citep{majumdar2023nsg}, which provides high-performance computing resources.

\subsection{Limitations}

This validation covers one 64-channel P300 dataset and only run~1. Resting-state, sleep, clinical, lower-density, and repeated-run data may expose other differences. We also tested only the default parameter set, which performed best in the original benchmark \citep{delorme2023eeg}. Other ICA algorithms, ICLabel thresholds, and filter settings remain untested. Nor did we measure the effect on time-frequency decomposition, source localization, or connectivity.

Within an architecture, no ICLabel probability lay close enough to 0.9 for the inference difference between Python and MATLAB to change a rejection. Other datasets may contain such components, and a probabilistic treatment of classification uncertainty may then be preferable. Architecture itself did change two decisions: sub-003 and sub-006 each lost two components on ARM but one on Intel, giving totals of 15 and 13. Python and MATLAB agreed within each architecture, but small differences between Apple Accelerate and Intel MKL reached ICA and moved those components across the threshold. Sub-microvolt data differences can therefore change a discrete decision near 0.9.

\section{Conclusion}

EEGPrep reproduced the MATLAB EEGLAB workflow to numerical precision. Within the workflow, ICLabel inference introduced the first measurable error, with RMS below $2.1 \times 10^{-5}$~µV after the full pipeline and no change in rejection decisions within either tested architecture.

EEGPrep is available from GitHub (\url{https://github.com/sccn/eegprep}), PyPI (\texttt{pip install eegprep}), and Docker Hub. The package reads BIDS data, writes BIDS derivatives, and is available through PyPI and Docker. These choices permit local or hosted processing without a MATLAB licence and make the software environment recordable. Users seeking exact replication should record the seed, package versions, and BLAS build.

\begin{samepage}
\section*{Acknowledgments}

This work was supported by NIH grants R01EB023297 and R24MH120037, which fund the Open EEGLAB Portal and NEMAR, respectively. We thank the NEMAR team for providing access to the BIDS datasets used in validation.
\end{samepage}

\section*{Conflict of interest}

The authors declare no competing interests. EEGPrep is released as free and open-source software.

\section*{Ethical statement}

This study used only the publicly available, fully de-identified BIDS dataset ds003061, obtained from the NEMAR repository. No new data were collected from human participants; accordingly, no additional ethical approval was required for this secondary analysis of openly available data.

\section*{Data and code availability}

EEGPrep is available at \url{https://github.com/sccn/eegprep} and via PyPI (\texttt{pip install eegprep}). Docker images are available at \url{https://hub.docker.com/r/arnodelorme/eegprep}. The validation dataset (ds003061) is publicly available through NEMAR (\url{https://nemar.org}). All analysis scripts and validation results are available in the project repository.

\bibliographystyle{iopart-num}
\bibliography{eegprep_bibliography}

\clearpage
\setcounter{table}{0}
\renewcommand{\tablename}{Supplementary Table}
\renewcommand{\theHtable}{S\arabic{table}}

\begin{table}[!htbp]
\centering
\caption{Stage-by-stage numerical parity on Intel x86\_64 macOS. We repeated the seven-stage analysis with Python 3.10, NumPy 1.x, and MATLAB R2022b. NumPy and MATLAB both linked Intel MKL, and Python used the native PyTorch ICLabel implementation. As in the primary ARM analysis, sub-012 was excluded, giving $N = 12$ for run~1. Values are in µV. All subject-level RMS values through Step~4 were approximately $10^{-12}$~µV, which was the same numerical-zero result found on ARM. Because NumPy and MATLAB both linked Intel MKL, re-reference and clean\_rawdata used the same floating-point accumulation order. PyTorch ICLabel inference introduced the first measurable error at Step~5, and RMS remained $\leq 2.4 \times 10^{-5}$~µV after interpolation and epoching. For all 12 subjects, the Picard AMARI distance was $\leq 0.000001$ and mean component correlation was 1.000, matching the primary analysis. Python and MATLAB made the same ICLabel decisions within each architecture. Intel rejected 13 components and ARM rejected 15 because sub-003 and sub-006 each lost one component on Intel and two on ARM. Both architectures used the same PyTorch ICLabel implementation. The cross-architecture difference therefore arose earlier, when small numerical differences between Intel MKL and Apple Accelerate passed through ICA and moved two near-threshold probabilities across 0.9. On ARM, Python used Apple Accelerate and MATLAB R2025a used a different linear-algebra library, yet their outputs still agreed through Picard to numerical precision. The result does not require a shared BLAS library, although it does require backends that resolve reductions compatibly.}
\label{tab:intel_parity_summary}
\begin{tabular}{lrccc}
\toprule
\textbf{Step} & $N$ & \textbf{Mean RMS} & \textbf{Max RMS} & \textbf{Max MaxAbsDiff} \\
\midrule
1. Import & 12 & 0 & 0 & 0 \\
2. Re-reference & 12 & 0 & 0 & 0 \\
3. clean\_rawdata & 12 & $1.3 \times 10^{-12}$ & $1.9 \times 10^{-12}$ & $2.3 \times 10^{-11}$ \\
4. Picard ICA & 12 & $1.3 \times 10^{-12}$ & $1.9 \times 10^{-12}$ & $2.3 \times 10^{-11}$ \\
5. ICLabel & 12 & $4.1 \times 10^{-6}$ & $2.2 \times 10^{-5}$ & $3.3 \times 10^{-4}$ \\
6. Interpolation & 12 & $4.1 \times 10^{-6}$ & $2.3 \times 10^{-5}$ & $3.3 \times 10^{-4}$ \\
7. Epoch & 12 & $4.1 \times 10^{-6}$ & $2.4 \times 10^{-5}$ & $3.0 \times 10^{-4}$ \\
\bottomrule
\end{tabular}
\end{table}

\clearpage
\begin{table}[h]
\centering
\caption{Picard ICA parity per subject (run~1): Python EEGPrep vs MATLAB eeglabcompat. AMARI distance and component correlation were computed on matched mixing matrices.}
\label{tab:ica_parity_json}
\begin{tabular}{lccccc}
\toprule
\textbf{Subject} & \textbf{Comp} & \textbf{AMARI distance} & \textbf{Corr (mean)} & \textbf{Corr (min)} \\
\midrule
sub-001 & 57 & 0.000000 & 1.000 & 1.000 \\
sub-002 & 33 & 0.000000 & 1.000 & 1.000 \\
sub-003 & 55 & 0.000000 & 1.000 & 1.000 \\
sub-004 & 56 & 0.000000 & 1.000 & 1.000 \\
sub-005 & 50 & 0.000001 & 1.000 & 1.000 \\
sub-006 & 52 & 0.000000 & 1.000 & 1.000 \\
sub-007 & 54 & 0.000000 & 1.000 & 1.000 \\
sub-008 & 59 & 0.000001 & 1.000 & 1.000 \\
sub-009 & 57 & 0.000000 & 1.000 & 1.000 \\
sub-010 & 56 & 0.000000 & 1.000 & 1.000 \\
sub-011 & 33 & 0.000000 & 1.000 & 1.000 \\
sub-013 & 60 & 0.000000 & 1.000 & 1.000 \\
\bottomrule
\end{tabular}
\end{table}

\clearpage
\begin{table}[h]
\centering
\caption{ICLabel component rejection per subject (run~1). Rejection threshold: Muscle~$\geq 0.9$ or Eye~$\geq 0.9$.}
\label{tab:iclabel_parity}
\begin{tabular}{lccc}
\toprule
\textbf{Subject} & \textbf{Rej (MAT)} & \textbf{Rej (PY)} & \textbf{Match} \\
\midrule
sub-001 & 2 & 2 & Yes \\
sub-002 & 0 & 0 & Yes \\
sub-003 & 2 & 2 & Yes \\
sub-004 & 0 & 0 & Yes \\
sub-005 & 3 & 3 & Yes \\
sub-006 & 2 & 2 & Yes \\
sub-007 & 0 & 0 & Yes \\
sub-008 & 2 & 2 & Yes \\
sub-009 & 2 & 2 & Yes \\
sub-010 & 0 & 0 & Yes \\
sub-011 & 1 & 1 & Yes \\
sub-013 & 1 & 1 & Yes \\
\bottomrule
\end{tabular}
\end{table}

\end{document}